\documentclass[%
 aip,
 amsmath,amssymb,
 reprint,%
]{revtex4-1}

\usepackage{graphicx}
\usepackage{dcolumn}
\usepackage{bm}

\usepackage[utf8]{inputenc}
\usepackage[T1]{fontenc}

\usepackage{mathptmx}
\usepackage{siunitx}
\usepackage{graphicx}
\usepackage{dcolumn}
\usepackage{braket}
\usepackage{xfrac}
\usepackage{hyperref}
\usepackage[usenames,dvipsnames]{color}
\usepackage{tikz-cd}
\usepackage{float}
\usepackage[nameinlink,capitalise]{cleveref}
\usepackage{mathtools}
\usepackage{amsmath}
\usepackage{bm}
\DeclareSIUnit\gauss{G}
\DeclareSIUnit{\au}{{a.u.}}
\hypersetup{
colorlinks=true,
linkcolor=blue,
citecolor=blue,
filecolor=green,
urlcolor=blue,
}

\begin{document}


\title{
Rainbow scattering in rotationally inelastic collisions of HCl and H$_2$
}

\author{Masato Morita}
\affiliation{Department of Chemistry and Biochemistry, University of Nevada, Las Vegas, Nevada 89154, USA}
\author{Junxiang Zuo}
\affiliation{Department of Chemistry and Chemical Biology, University of New Mexico, Albuquerque, New Mexico 87131, USA}
\author{Hua Guo}
\affiliation{Department of Chemistry and Chemical Biology, University of New Mexico, Albuquerque, New Mexico 87131, USA}
\author{Naduvalath Balakrishnan}
\affiliation{Department of Chemistry and Biochemistry, University of Nevada, Las Vegas, Nevada 89154, USA}


\begin{abstract}
We examine rotational transitions of HCl in collisions with H$_2$ by carrying out  quantum mechanical close-coupling and quasi-classical trajectory calculations on a recently developed globally accurate  full-dimensional  {\it ab initio} potential energy surface for the H$_3$Cl system.
Signatures of rainbow scattering in rotationally inelastic collisions are found in the state resolved integral and differential cross sections as functions of the impact parameter (initial orbital angular momentum) and final rotational quantum number.
We show the coexistence of distinct dynamical regimes for the HCl rotational transition driven by the short-range repulsive and long-range attractive forces whose relative importance depends on the collision energy and final rotational state suggesting that classification of rainbow scattering into rotational and $l$-type rainbows is effective for H$_2$+HCl collisions. 
While the quasi-classical trajectory method satisfactorily predicts the overall behavior of the rotationally inelastic cross sections, its capability to accurately describe signatures of rainbow scattering appears to be limited for the present system. \end{abstract}
\maketitle

\section{Introduction}
\label{sec:Introduction}

Molecular collision and chemical reaction outcomes are strongly influenced by initial internal states of reactants, collision energy, kinematics, stereodynamics, intermolecular forces, external fields, etc.
Identifying the collisional parameters that control the collision outcomes has received much attention in theoretical and experimental studies.~\cite{levine2009molecular,child_1996,LevinBernstein,ShinkeBowman,Forrey_ML}
The Wigner threshold law,~\cite{Wigner} Breit-Wigner and Fano-Feshbach profiles for resonances,~\cite{Breit_Wigner,Feshbach_1,Feshbach_2,Fano} and the Langevin model~\cite{Langevin,Cote_Dalgarno} are well-known examples of useful concepts to characterize the physical origin and behavior of cross sections as a function of collision energy.
In the rapidly evolving field of ultracold atoms and molecules, zero-energy Feshbach resonances as functions of external magnetic and/or electric fields play a central role in the control of the collision outcome.~\cite{Krems_res,Julienne_res1,Julienne_res2}
Cross sections and rate coefficients for barrierless reactions and inelastic collisions are controlled by the values of long-range coefficients such as $C_4$ and $C_6$ in capture models from the ultracold to the Langevin energy regime.~\cite{levine2009molecular,ion_Willitsch} 

A long sought goal in molecular dynamics is to predict parameter dependence of scattering properties from limited information. 
An extreme example is the $J$-shifting approximation~\cite{J-shift} for predicting the behavior of reaction probabilities for non-zero values of the total angular momentum quantum number $J$ from that of $J=0$. While this approach has found some success for activated reactions it does not work well for complex-forming reactions and inelastic scattering.
More recently, novel methods based on machine learning are actively being developed to interpolate cross sections and rate coefficients as  functions of energy and temperature.~\cite{Krems_PCCP_2019,Koner_2019,Bowman_ML_1,Bowman_ML_2,Forrey_ML,Chen2020}

Rainbow scattering, leading to characteristic parameter dependence of the scattering properties, may occur in systems  with isotropic potentials and multiple partial waves.~\cite{levine2009molecular,child_1996,Schinke_l_Li+-H2,Bowman_1979,Schinke_1979,Thomas_rev,Schinke_1981,Schinke_1982,ShinkeBowman}
Rainbow features arise from an interplay between the short-range repulsive and long-range attractive forces.
In classical mechanics, a signature of rainbow is observed as a singularity (divergence and discontinuity) in the elastic differential cross section (DCS) at the maximum negative deflection angle  $\chi(b)$ with the impact parameter $b=b_\text{r}$ satisfying $(d\chi(b)/db)_{b_\text{r}}=0$, where the relation between deflection angle and scattering angle is $|\chi(b)|=\theta$.~\cite{levine2009molecular,LevinBernstein,child_1996}
Semiclassically,  rainbow signature in the DCS is an oscillatory quantum interference pattern associated with a confluence of two paths scattered into the same $\theta$ with different values of $b$ or orbital angular momentum quantum number $l$ ($\approx bk$, where $k$ is the wave vector) due to the condition $(d\chi(b)/db)_{b_\text{r}}=0$ or $(d\chi(l)/dl)_{l_\text{r}}=0$.~\cite{FordWheeler,levine2009molecular,child_1996,ShinkeBowman} 
The sensitivity of the rainbow signature in the DCS to small changes in the interaction potential can shed insights into the improvement of the potential used in the calculation.~\cite{levine2009molecular,child_1996,Bowman_1979,Thomas_rev}

Rainbow scattering in the DCSs for rotationally inelastic ($j \to j'$) collisions in atom+molecule and ion+molecule systems has been the subject of both experimental and theoretical investigations.~\cite{Bowman_1979,Schinke_1979,Thomas_rev,Schinke_1981,Schinke_1982,ShinkeBowman}
Signatures of rainbow scattering can be characterized by the excitation function $j'(l,\gamma)$ and the deflection function $\chi(l,\gamma)$ where $\gamma$ is the molecular orientation angle against the initial relative velocity vector for the collision.
Classically, a singularity in the DCS appears when the determinant $D(l,\gamma)=(\partial \chi/\partial l)_\gamma (\partial j'/\partial \gamma)_l - (\partial \chi/\partial \gamma)_l (\partial j'/\partial l)_\gamma$  of the Jacobian  of the dynamical mapping between ($l, \gamma$) and ($\chi,j'$) becomes zero.  
Semiclassically, quantum interference pattern is observed when "adjacent" trajectories with slightly different initial values of $l$ and $\gamma$ give rise to the same final $J$ and $\chi$.
For impulsive collisions, $D(l,\gamma)$ may be well approximated by the diagonal terms as $D(l,\gamma)=(\partial \chi/\partial l)_\gamma (\partial j'/\partial \gamma)_l$, suggesting that the singularity is classified according to whether $(\partial \chi/\partial l)_\gamma$ or $(\partial j'/\partial \gamma)_l$ vanishes.~\cite{Bowman_1979,Schinke_1979,Thomas_rev,Bowman_1981,Schinke_1981,Schinke_1982} 
When $(\partial \chi/\partial l)_\gamma=0$ the rainbow is called an $l$-type (or potential or impact parameter) rainbow, similar to that observed in elastic scattering.
Another type is rotational (or orientation or stereodynamic) rainbow which can occur due to the inner repulsive region even in the absence of a potential well.
In contrast to elastic scattering, rotationally inelastic scattering requires anisotropy in the potential to change the rotational level. 
While the anisotropy of the potential is usually large enough at the short-range to cause rotational transitions (provided energetically allowed), the anisotropy in the long-range attractive part is not necessarily strong enough, in particular for neutral species. 
Thus, $l$-type rainbow is generally prominent only for small changes in rotational quantum numbers. 
Indeed, many of the early studies have focused on rotational rainbow scattering. 
Aoiz and cowokers have examined the role of long-range attractive forces in $l$-type rainbow in a series of studies of rotationally inelastic atom\,+\,diatomic molecule collisions:  Ar+NO, Cl+H$_2$, Ne+NO, Xe+NO, and Kr+NO.~\cite{Aoiz_2003_Ar-NO,Aoiz_2011_Cl-H2,Aoiz_2012_Xe-NO,Aoiz_2014_Kr-NO} 
We note that there is no definitive criterion to classify the rainbows into the two types in advance.~\cite{Thomas_rev,Bowman_1981,Schinke_1982} 

Recent experimental progress in high-resolution measurements of state-to-state DCSs and stereodynamic control of rotationally inelastic collisions may facilitate observation of rainbows with fewer averaging effects.~\cite{Aoiz_2014_Kr-NO,Onvlee_2015,2017_Science_Perreault,DCS_McKendrick,DCS_Suit,DCS_Meerakker}
Indeed, rainbow structures have been observed in the state-to-state DCS of rotationally inelastic collisions for more complex systems such as He+H$_2$O,~\cite{H2O-He} H$_2$+NO,~\cite{NO-H2_2017} D$_2$+NO,~\cite{NO-H2_2017,Stole_2006} and CH$_4$+NO~\cite{Orr-Ewing_NO-CH4}.
Furthermore, signatures  of {\it reactive rainbows} have been observed in reactions F+CH$_3$D $\to$ CH$_2$D + HF and F+CH$_3$D $\to$ CH$_3$ + DF by Liu and coworkers.~\cite{Kopin_2016,Kopin_2020}
On the other hand, full-dimensional quantum scattering calculations for exploring rotational rainbows are rare for molecule\,+\,molecule collisions.  
Recently, within a reduced four-dimensional (4D) treatment,  detailed quantum scattering calculation were reported for the DCS in H$_2$+NO collisions.~\cite{NO-H2_2017} 
In addition, even for atom\,+\,diatomic molecule collisions, full-dimensional quantum scattering calculations are scant for open-shell molecules regardless of the appearance of rainbow scattering except for very recent reports on Ar+NO and H+NO collisions~\cite{NO-Ar_full}.

In this paper, we report explicit six-dimensional (6D) quantum mechanical scattering calculations and quasi-classical trajectory (QCT) calculations of integral and differential cross sections for rotational excitation of HCl by {\it para}-H$_2$ at collision energies ranging from 
$E_\text{c}=100$ cm$^{-1}$ to $6000$ cm$^{-1}$. 
Quantum calculations of rotational relaxations in HCl by H$_2$ within a 4D rigid rotor model were previously reported by Lanza and coworkers.~\cite{Lanza_PES} 
In our recent work we have reported a 6D potential energy surface (PES) for the H$_2$+HCl system~\cite{H2HCl_PES} as well as 6D quantum calculations of rotational transitions in cold collisions with H$_2$ with an emphasis on stereodynamic control at around 1K~\cite{Morita_HCl}.
Unlike rare gas-molecule collisions, the interaction potential between HCl and H$_2$ is much more anisotropic making it a more compelling case for rainbow scattering in rotationally inelastic molecule+molecule collisions. Whether the lightness of the collision partners and the high asymmetry of the masses of Cl and H may preclude rainbow scattering in this system is an open question.
While the experimental DCS for the rotational excitation of HCl by heavier colliders, e.g., rare gas atoms (Ne, Ar and Kr), N$_2$ and CH$_4$, was reported by Chandler and coworkers, the energy range and the number of final rotational states explored are limited.~\cite{Chandler_PCCP,Chandler_JPCA}

\section{Theory and computation}
\label{sec:Theory and computation}

\subsection{Potential energy surface}
\label{subsec:PES}

The computations  are performed using a recently developed globally accurate 6D {\it ab initio} PES for the electronic ground state of the H$_3$Cl system~\cite{H2HCl_PES}.
Briefly, the PES was fitted using a permutation-invariant polynomial neural network (PIP-NN) technique \cite{PIP-NN_1,PIP-NN_2} based on a grid of interaction energies calculated with the CCSD(T)-F12b \cite{F12b}/aug-cc-pVQZ \cite{AVQZ} method. 
The long-range part of the interaction potential between H$_2$ and HCl includes an accurate description of the electrostatic and dispersion interactions. 
The well-depth of the interaction potential at the global minimum is -215.5 cm$^{-1}$ corresponding to a T-shaped geometry with H$_2$ molecule on the H side of the HCl molecule. A second  minimum with a depth of -102.6 cm$^{-1}$ exists for another T-shaped structure with H$_2$ on the Cl side of HCl. 
These  potential wells are significantly deeper than that between two H$_2$ molecules (-35 cm$^{-1}$). 
At the global minimum, the distance between the centers of mass of H$_2$ and HCl  is $R=3.55$ \AA.
In our previous work~\cite{H2HCl_PES} we have compared key features of this PES with the 4D PES of Lanza and coworkers.~\cite{Lanza_PES} 
and confirmed its accuracy for pure rotational transitions in HCl induced by H$_2$.

\subsection{Quantum scattering calculation}
\label{subsec:CC}

We numerically solve the time-independent Schr\"{o}dinger equation for H$_2$+HCl collisions based on the close-coupling (CC) method~\cite{1960_Dalgarno_Arthurs} in the space-fixed (SF) coordinate frame within the total angular momentum representation for the angular degrees of freedom as implemented in a modified version of the TwoBC code. \cite{TwoBC,H2H2_CC}
The computational details, including relevant basis set parameters are discussed in our previous publications,~\cite{H2HCl_PES,Morita_HCl} thus we provide only a brief outline of the scattering formalism.

The Hamiltonian for the relative motion of two diatomic molecules ($^{1}\Sigma$) in a set of Jacobi vectors ($\bm{r}_1, \bm{r}_2, \bm{R}$) in the SF coordinate frame is given by ($\hbar=1$ hereafter)
\begin{equation}
{
\hat{\mathcal{H}} = 
- \frac{1}{2\mu R} \frac{d^2}{dR^2}R 
 + \frac{\hat{\bm{l}}^2}{2\mu R^2}
 + \hat{H}_\text{asym}
 + U_\text{int}(\bm{R},\bm{r}_1,\bm{r}_2)
},
\label{eq:Heff}
\end{equation}
where $\mu$ is the reduced mass of the collision partners, $R$ is the magnitude of the vector $\bm{R}$ joining the centers of mass of the two molecules, $\hat{\bm{l}}$ is the orbital angular momentum operator for the relative motion, 
$U_\text{int}(\bm{R},\bm{r}_1,\bm{r}_2)$ is the interaction potential, and $\hat{H}_\text{asym}$ is the asymptotic Hamiltonian expressed as the sum of the Hamiltonians of the separated molecules, $\hat{H}_\text{asym} = \hat{h}_\text{1}(\bm{r}_1)+\hat{h}_\text{2}(\bm{r}_2)$. 
The Hamiltonian for each of the diatomic molecule, $\hat{h}_i(\bm{r}_i)$ ($i=1,2$) is 
\begin{equation}
{
\hat{h}_i(\bm{r}_i) = - \frac{1}{2\mu_i r_i} \frac{d^2}{dr_i^2}r_i + \frac{{\hat{\bm{j}_i}^2}}{2\mu_i {r_i}^2}+V_i(r_i)
},
\label{eq:Hmol}
\end{equation}
where $\mu_i$ is the reduced mass, $\hat{\bm{j}_i}$ is rotational angular momentum operator, and $V_i(r_i)$ is the potential energy function of the diatom.
The  PES  of the tetra-atomic system, $V(\bm{R},\bm{r}_1,\bm{r}_2)$ is expressed as
\begin{equation}
{ V(\bm{R},\bm{r}_1,\bm{r}_2)=V_1(r_1)+V_2(r_2)+U_\text{int}(\bm{R},\bm{r}_1,\bm{r}_2)}. 
\label{eq:PES}
\end{equation}

Since the total angular momentum {$\bm J$} of the collision complex, its projection $M$ onto the SF z-axis, and the inversion parity $\epsilon_I=(-1)^{j_1+j_2+l}$ are conserved, the total wavefunction for given values of $J,~M$, and $\epsilon_I$ may be expanded as 
\begin{equation}
\label{eq:basis}
\Psi^{JM\epsilon_I} = \frac{1}{R} \sum^{}_{v\,j\,l}
F^{JM\epsilon_I}_{vjl}(R) \frac{\phi^{j_1}_{v_1}(r_1)}{r_1} \frac{\chi^{j_2}_{v_2}(r_2)}{r_2}|JM\epsilon_I(lj_{12}(j_1j_2))>,
\end{equation}
where $v=v_1v_2$ and $j=j_1j_2j_{12}$ denote the vibrational and rotational quantum numbers collectively, $F^{JM\epsilon_I}_{vjl}(R)$ are the radial expansion coefficients in $R$, 
$\phi^{j_1}_{v_1}(r_1)/r_1$ and $\chi^{j_2}_{v_2}(r_2)/r_2$ denote radial parts of the rovibrational eigenfunctions of the diatomic molecules with the Hamiltonian $\hat{h}_i(\bm{r}_i)$ in \cref{eq:Hmol}, and $|JM\epsilon_I(lj_{12}(j_1j_2))>$ denotes eigenstates of the total angular momentum for a given  $M$ and parity $\epsilon_I$.

The coefficients $F^{JM\epsilon_I}_{vjl}(R)$ satisfy the CC equations obtained by substituting \cref{eq:basis} into the Schr\"{o}dinger equation,
\begin{widetext}
\begin{align}
\Bigl{ [ }  & \frac{1}{2\mu } \frac{d^2}{dR^2}\  - \frac{l(l+1)}{2\mu R^2} + E_\text{c} \Bigr{ ] } F^{JM\epsilon_I}_{vjl}(R) \nonumber \\
&= \sum^{}_{v'\,j'\,l'} F^{JM\epsilon_I}_{v'j'l'}(R) 
\int_0^{\infty} \int_0^{\infty} \langle JM\epsilon_I(lj_{12}(j_1j_2) | 
\phi^{j_1}_{v_1}(r_1) \chi^{j_2}_{v_2}(r_2) 
  U_\text{int}(\bm{R},\bm{r}_1,\bm{r}_2)
\phi^{j^\prime_1}_{v^\prime_1}(r_1) \chi^{j^\prime_2}_{v^\prime_2}(r_2) |JM\epsilon_I(l'j'_{12}(j'_1j'_2)\rangle dr_1 dr_2.
\label{eq:CC}
\end{align}
\end{widetext}
For a given total energy $E$, the collision energy
$E_\text{c}=E-E_{v_1j_1}-E_{v_2j_2}$ where  $E_{v_ij_i}\,(i=1,2)$ denote the initial ro-vibrational energies for the diatomic molecules. 
Note that the CC equations in \cref{eq:CC} are independent of $M$.

In the TwoBC code, the matrix elements of the interaction potential given in the right hand side of \cref{eq:CC} are evaluated by expanding the angular dependence ($\hat{\bm{R}},\hat{\bm{r}}_1,\hat{\bm{r}}_2$) of the interaction potential in a triple series of spherical harmonics~\cite{Green_JCP_1975}.
A Gauss-Hermite quadrature is used to evaluate the integrals over  $r_1$ and $r_2$ of the expansion coefficients of the interaction potential with the vibrational wave functions $\phi(r_1)$ and $\chi(r_2)$. 
The explicit form of the matrix elements and relevant details are given in Ref. [\onlinecite{H2H2_CC}].
The radial propagation of the CC equations is carried out by Johnson's log-derivative propagation method and the S-matrix is obtained by  applying asymptotic boundary conditions at $R_\text{max}=50.0$ \AA\ at each energy.  
The S-matrix carries the relevant information to compute the state-resolved differential and integral cross sections. 

In the following, we omit the symbol $M$ and $\epsilon_I$ for simplicity. 
The state-to-state differential cross section (DCS) may be written in terms of the scattering amplitude $q$ in the helicity representation as \cite{DCS,2019_JCP_Croft}
\begin{equation}
\label{eq:DCS_k}
\small
\begin{split}
     \frac{ d\sigma_{{\alpha} \to \alpha'} } {d\Omega} 
    = \frac{1}{(2j_1+1)(2j_2+1)} 
     \sum^{}_{k_1,k_2,k'_1,k'_2}
  |q_{\alpha,k_1,k_2 \to 
      \alpha', k'_1,k'_2}|^2,
\end{split}
\normalsize
\end{equation}
where $\alpha\equiv v_1j_1v_2j_2$ and $\alpha'\equiv v'_1j'_1v'_2j'_2$ refer to the initial and final combined molecular rovibrational states, respectively, $d\Omega$ is the infinitesimal solid angle,
the quantum number $k$/$k'$ is the helicity component of the initial/final molecular rotation angular momentum $j/j'$. 
The contributions of all possible degenerate $k$/$k'$ components in $j/j'$ are summed over in the right hand side of \cref{eq:DCS_k}. 

The  scattering amplitude $q$ is given by \cite{DCS}
\begin{equation}\label{eq:q_k}
\small
\begin{split}
& q_{\alpha,k_1,k_2  \to  \alpha', k'_1,k'_2} \\
&= \frac{1}{2 k_{\alpha}} 
      \sum^{}_{J} (2J+1)  
 \sum^{}_{j_{12}, j'_{12}, l, l'} i^{l-l'+1}  T^{J}_{\alpha j_{12} l,\alpha' j'_{12} l'}(E) d^{J}_{k_{12},k'_{12}}(\theta) \\
&\ \ \ \  \times <j'_{12} k'_{12} J -k'_{12}|l' 0> <j_{12} k_{12} J -k_{12}|l 0> \\
&\ \ \ \  \times <j'_{1} k'_{1}   j'_{2} k'_{2}|j'_{12} k'_{12}> <j_{1} k_{1} j_{2} k_{2}|j_{12} k_{12}>,
\end{split}
\normalsize
\end{equation}
where $k_{\alpha}$ is the wave vector for the incident channel, the T-matrix is given in terms of the S-matrix as $T^J(E)=1-S^J(E)$, the projection quantum numbers $k_{12}=k_1 + k_2$ and $k'_{12}=k'_1 + k'_2$,
 $d^{J}_{k_{12},k'_{12}}(\theta)$ is the Wigner's reduced rotation matrix, and the braket $< \ | >$ denotes a Clebsch-Gordan coefficient. 
The scattering amplitude includes  energy dependence through $k_{\alpha}$ and the T-matrix, and angle dependence through $d^{J}_{k_{12},k'_{12}}(\theta)$. 
By taking the integral of the DCS (\cref{eq:DCS_k}) over $\theta$ (0 to $\pi$) and $\phi$ (0 to $2\pi$), one obtains the state-to-state integral cross section (ICS) as \cite{DCS,2019_JCP_Croft,H2HCl_PES}
\begin{align}
\label{eq:ICS}
&\sigma_{{\alpha} \to \alpha'} (E) \nonumber \\
 &=\frac{\pi}{(2j_1+1)(2j_2+1)k_{\alpha}^2}  \sum^{}_{j_{12}j'_{12},ll'J}(2J+1) |T^{J}_{\alpha j_{12} l,\alpha' j'_{12} l'}(E)|^2.
\end{align}

\begin{figure*}[t!]
\begin{center}
\includegraphics[scale=0.36]{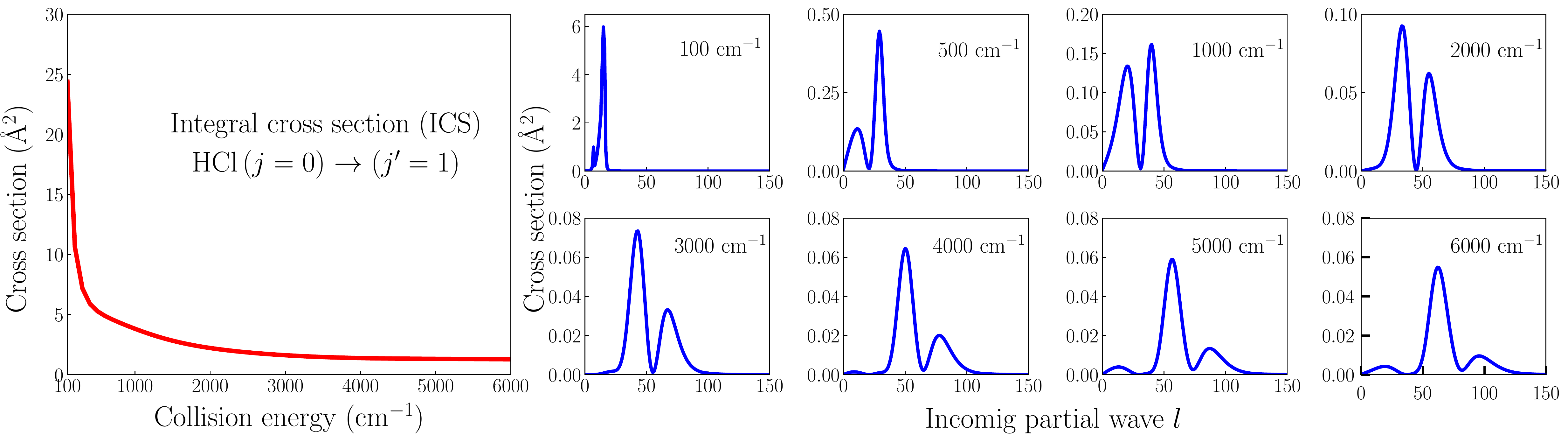}
\end{center}
\vspace{-5mm}
\caption{
Integral cross section for HCl ($v=0, j=0$) + H$_2$ ($j=0$) $\to$ HCl ($v'=0, j'=1$) + H$_2$ ($j'=0$) rotationally inelastic collisions at energies ranging from $E_\text{c}=100$ cm$^{-1}$ up to $6000$ cm$^{-1}$ (left panel).
Partial cross section decomposition with respect to the incoming partial wave $l$ are displayed in the right panels at  $E_\text{c}=100, 500, 1000, 2000, 3000, 4000, 5000$ and $6000$ cm$^{-1}$. 
Results are from quantum mechanical close-coupling calculations, thus $l$ is an integer. }
\label{fig:l-ICS}
\end{figure*}

\subsection{Quasi-classical trajectory method}
\label{subsec:QCT} 

The QCT calculations are carried out using the VENUS program package.~\cite{VENUS} 
Batches of $5 \times10^{5}$ trajectories were calculated at each collision energy on the 6D PES.
The maximum impact parameters, $b_\text{max}$, was determined to be 8 \AA\, after carrying out a small set of trajectories with trial values. 
All trajectories in QCT calculations were started from a diatom-diatom distance of 15 \AA\, and ended when the products or reactants reach a separation of 15.5 \AA.
The propagation time step was chosen to be 0.1 fs, which guarantees the energy conservation of all trajectories better than 0.01 kcal/mol ($\sim 3.5\ \text{cm}^{-1}$). 

The state-to-state ICS corresponding to \cref{eq:ICS} is calculated by 
\begin{equation}
{
\sigma_{\alpha \to \alpha'}(E)\ =\ \pi b^2_\text{max} \frac{N_{\alpha \to \alpha'}}{N_{\text{tot},\alpha}} 
},
\label{eq:QCT_ICS}
\end{equation}
where ${N_{\alpha \to \alpha'}}$ denotes the number of trajectories which result in a final combined molecular rovibrational state $\alpha'$ with an initial combined state $\alpha$. 
The umber of total trajectories with the initial combined state $\alpha$ is denoted as $N_{\text{tot},\alpha}$. 
In the present study, the initial combined state $\alpha$ is composed of the rovibrational ground states of HCl and H$_2$.
The values of the quantum numbers for the final rovibrational states were rounded to the nearest integers. 

The DCS is computed by 
\begin{equation}
{
\frac{d \sigma_{\alpha \to \alpha'}(E)}{d \Omega}\ =\ \frac{\sigma_{\alpha \to \alpha'}(E) P_{\alpha \to \alpha'}(\theta)}{2\pi \text{sin} \theta} 
},
\label{eq:QCT_DCS}
\end{equation}
where $\theta$ is the scattering angle, $P_{\alpha \to \alpha'}(\theta)$ is the normalized probability obtained by 
\begin{equation}
{
P_{\alpha \to \alpha'}(\theta)\ =\ \frac{\displaystyle \sum_{\theta' \geq \theta-\Delta \theta}^{\theta'< \theta+\Delta \theta}\ N_{\alpha \to \alpha'}(\theta')} {N_{\alpha \to \alpha'}},
}
\label{eq:QCT_Pr}
\end{equation}
where $N_{\alpha \to \alpha'}(\theta')$ is the number of trajectories in a bin from $\theta-\Delta \theta$ to $\theta+\Delta \theta$. In the computations, $\Delta \theta$ is taken to be $1.0 ^\circ$.

\section{Results}
\label{sec:Results}

We consider excitations of HCl from the rovibrational ground state ($v=0, j=0$) to the rotationally excited state ($v=0, j'=1$) in collisions with {\it para}-H$_2$ ($v=0, j=0$), namely HCl ($v=0, j=0$) + H$_2$ ($v=0, j=0$) $\to$ HCl ($v'=0, j'=1$) + H$_2$ ($v'=0, j'=0$), at collision energies ranging from $E_\text{c}=100$ cm$^{-1}$ to $6000$ cm$^{-1}$. 
At these collision energies, the vibrationally inelastic scattering is either energetically disallowed or very weak even when energy is sufficiently large, and we will  focus on  pure rotationally inelastic transitions of HCl. For the same reason, in subsequent discussions, we will suppress vibrational quantum numbers of both molecules.
While  low energy rotational transitions of HCl by H$_2$ exhibit isolated shape resonances~\cite{H2HCl_PES,Morita_HCl}  the resonances vanish as the collision energy is increased. Indeed,
the left panel of \cref{fig:l-ICS} shows ICS as a function of the collision energy which depicts a smooth behavior due to contributions from many  partial waves. 
However, when the ICS is examined as a function of $l$, $l'$ or $J$, we find multiple peaks and associated minima that dip to almost zero as  collision energy is increased. 
In the higher energy region ($E_\text{c} > 500$ cm$^{-1}$) where  many incoming and outgoing partial waves contribute the plots of ICS vs.~$l$, $l'$ and $J$ all look very similar. In particular, since both HCl and H$_2$ are initially in their rotational ground states ($j=0$), $l=J$ is satisfied.  
For subsequent analysis and discussions of the impact parameter dependence of the cross sections, we show the ICS as a function of the incoming partial wave $l$ in the right panels of \cref{fig:l-ICS} at selected values of collision energies, $E_\text{c}=100, 500, 1000, 2000, 3000, 4000, 5000,$ and $6000$ cm$^{-1}$.  
At each of these collision energies, the partial ICS summed over all integer $l$ yields the total ICS shown in the left panel. 
For energies between $E_\text{c}=500-3000$ cm$^{-1}$ the partial ICS vs. $l$ exhibits a double-peak structure and a third peak begin to  appear  in the lower $l$ region ($l<25$) for $E_\text{c}>3000$ cm$^{-1}$. 
The peak positions and minima shift to higher values of $l$ as the collision energy is increased. 
Also, the intensity of the peak at higher $l$ values decreases monotonically with  collision energy, indicating its diminishing importance compared to the dominant peak at lower $l$. 
More detailed analysis of the partial wave contributions and relation with elastic scattering are given in Supplementary Material (SM). 

\begin{figure}[t!]
\begin{center}
\includegraphics[scale=0.4]{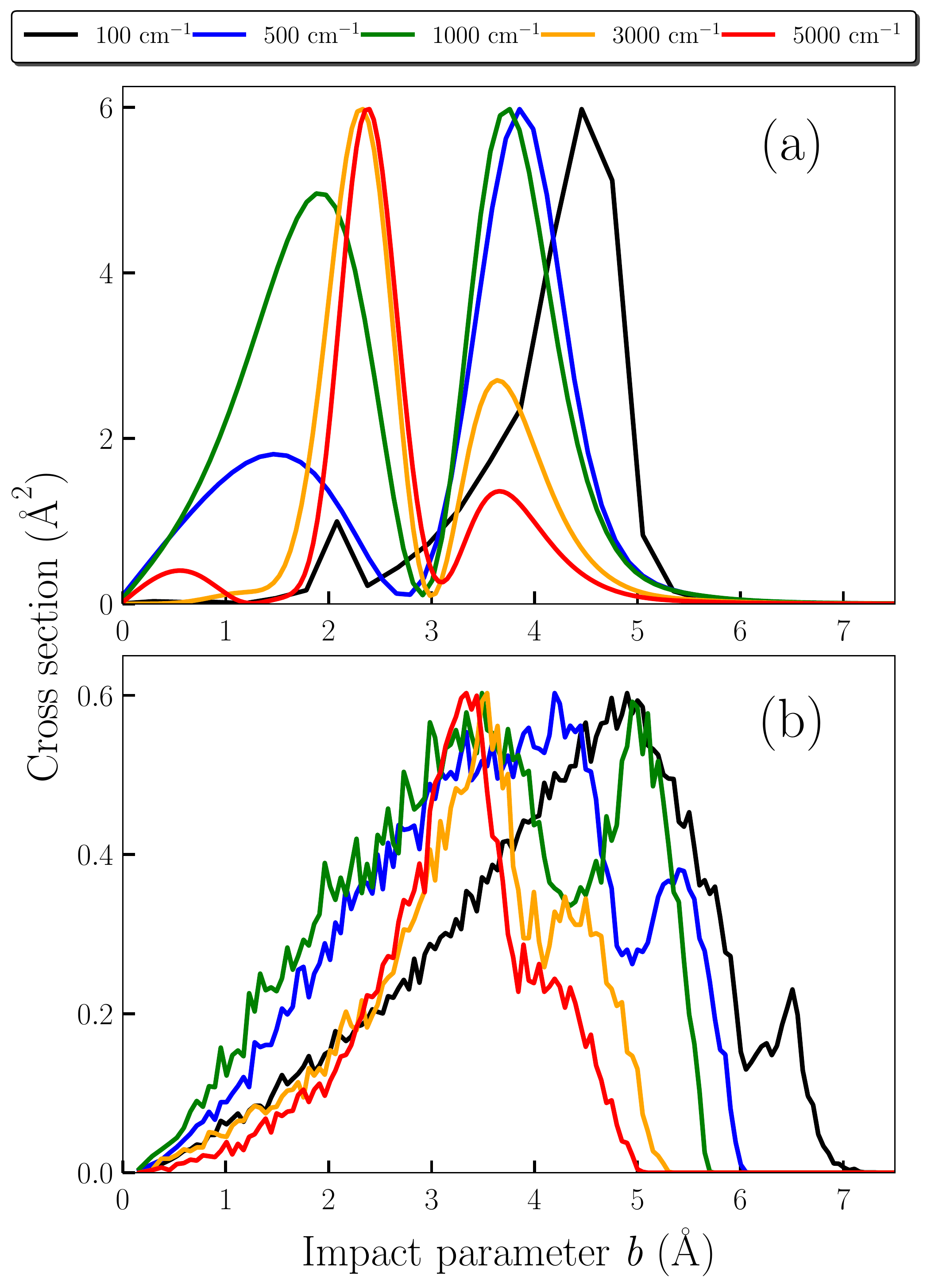}
\end{center}
\vspace{-5mm}
\caption{
Partial rotational inelastic cross section as a function of the impact parameter $b$ for HCl ($v=0, j=0$) to ($v'=0, j'=1$) in collisions with {\it para}-H$_2$ ($ j=0$).
(a) CC and (b) QCT. 
The partial cross section at each energy is normalized so that the maximum value becomes equal to the maximum at $E_\text{c}=100$ cm$^{-1}$ (black curve).
}
\label{fig:b-ICS}
\end{figure}

\begin{figure}[b!]
\begin{center}
\includegraphics[scale=0.33]{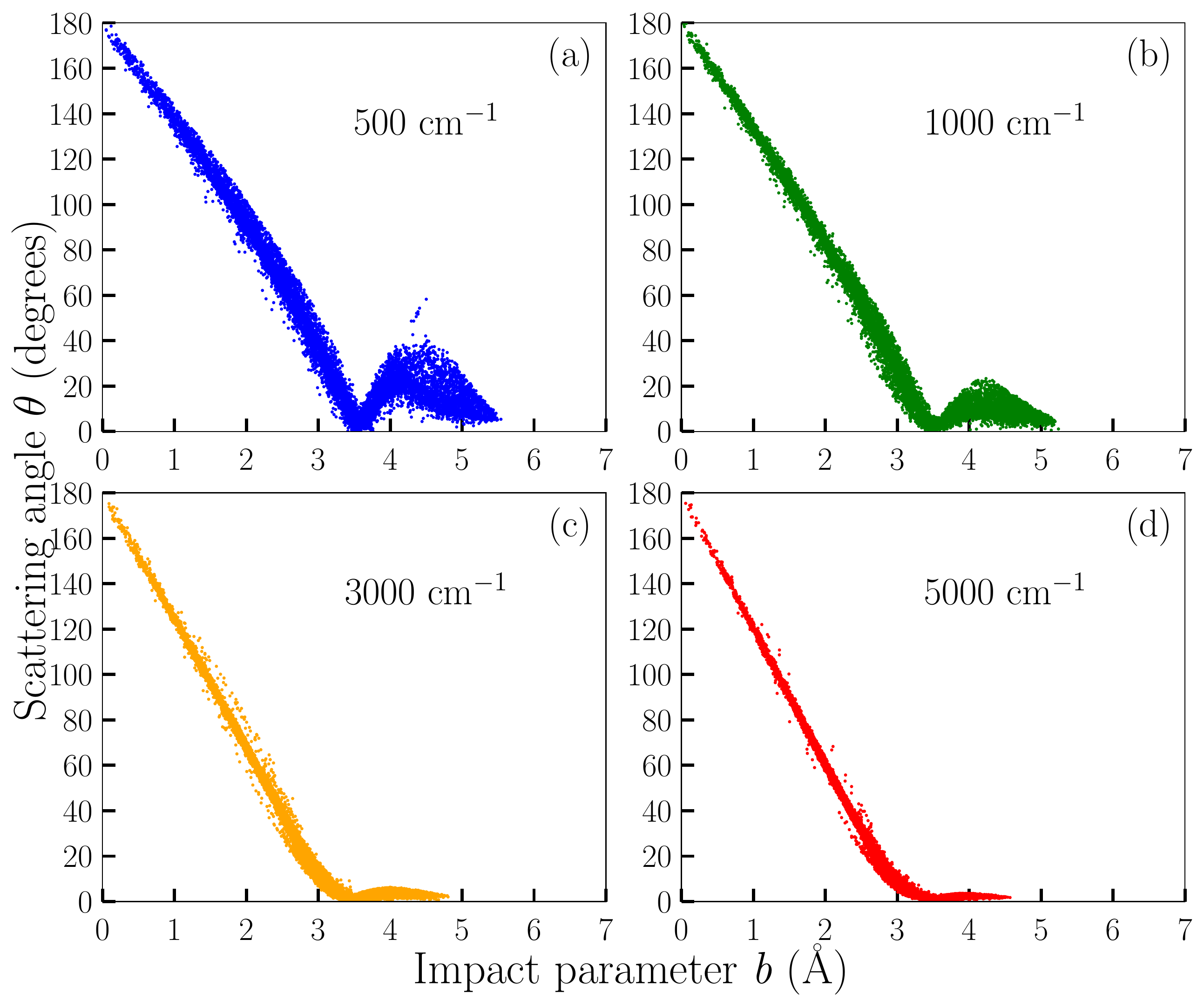}
\end{center}
\vspace{-5mm}
\caption{
Correlation diagrams between impact parameter $b$ and scattering angle $\theta$ with QCT for HCl ($v=0, j=0$) + H$_2$ ($j=0$) $\to$ HCl ($v'=0, j'=1$) + H$_2$ ($ j'=0$).
}
\label{fig:deflection}
\end{figure}

To gain more insights into the multi-peak structure of the partial ICS with respect to $l$ and to make a comparison with QCT results, we examine the impact parameter ($b$) dependence of the ICS plotted in \cref{fig:l-ICS}.
In the classical limit $b$ is given in terms of $l$ as $b=l/\sqrt{2\mu E_\text{c}}$. 
While the conversion between $l$ and $b$ is not unique due to the quantum nature of $\hat{\bm{l}}$, the non-uniqueness does not cause any issue in the following discussion since many partial waves contribute. 
The $b$ dependence of the partial ICS obtained from the results in \cref{fig:l-ICS} is shown in \cref{fig:b-ICS}~(a) at selected collision energies. 
It is clearly seen that the peaks and minima of the partial ICS depend weakly on the collision energy, thus we observe a minimum at around $b=3$ \AA \ and maximum on both sides of it regardless of the energy. 
Strictly, as the collision energy increases, the location of the minimum also increases from $b=2.86$ \AA \ at  $E_\text{c}=500$ cm$^{-1}$ to $b=3.13$ \AA \ at  $E_\text{c}=5000$ cm$^{-1}$.
At the lowest collision energy of $E_\text{c}=100$ cm$^{-1}$ the ICS is dominated by the peak at the right of the minimum. 
The relative importance of the peak at lower $b$ grows with the collision energy, and eventually the low-$b$ peak becomes dominant for collision energies above 3000 cm$^{-1}$.  
Furthermore, at the highest collision energy ($E_\text{c}=5000$ cm$^{-1}$), the low-$b$ peak is split into two distinct peaks with a minimum at around $b=1.5$ \AA.
The multi-peak structure, similar to \cref{fig:b-ICS}~(a), is also observed for
non-zero initial rotational level of HCl  and for DCl+H$_2$ collisions  (see SM). 

Generally, low impact parameters lead to head-on collisions and  backward scattering while high-$b$ approaches favor glancing collisions and forward scattering.  
Thus, \cref{fig:b-ICS}~(a) reveals that there are two competing collision mechanisms that correlate with low-$b$ and high-$b$ centered around $b\sim3$ \AA. 
Such clear separation of dynamical regimes by a minimum that becomes almost zero in the partial ICS or opacity function has been reported for ion+molecule and atom+molecule systems.~\cite{Schinke_l_Li+-H2,Barret_l,Aoiz_2003_Ar-NO,Aoiz_2011_Cl-H2,Aoiz_2012_Xe-NO,Aoiz_2019_H-HF} 
If the classification of rainbows into rotational and $l$-type rainbows is valid for H$_2$+HCl collisions, then the high-$b$ peak ($\sim3.8$ \AA) corresponds to an $l$-type rainbow.~\cite{Aoiz_2003_Ar-NO,Aoiz_2011_Cl-H2,Aoiz_2012_Xe-NO}
The relative dominance of the high-$b$ peak at lower collision energies in  \cref{fig:b-ICS}~(a) implies that the scattering is dominated by long-range force, thus the energy dependence of the high-$b$ peak is consistent with the character of an $l$-type rainbow. 
The rainbow impact parameter $b_\text{r}$ for the potential ($l$-type rainbow) is roughly approximated by the location of the potential minimum~\cite{levine2009molecular}. 
As stated in~\cref{subsec:PES}, the minimum of the potential corresponds to $R=3.55$ \AA, thus the position of the high-$b$ peak can be reasonably approximated  as an $l$-type rainbow.  
  
Before delving further into the analysis of the high-$b$ peak, we consider the complementary QCT results depicted in \cref{fig:b-ICS}~(b).  
Overall, the behavior of the QCT results as a function of $b$ exhibits similar energy dependence to that shown in \cref{fig:b-ICS}~(a). 
However, the clear modal structure obtained with the CC calculation is not reproduced by QCT. 
We can identify that the high-$b$ peaks are located at $b=4.9$ \AA~ and $b=4.6$ \AA\ at $E_\text{c}=500$ cm$^{-1}$ (blue) and $1000$ cm$^{-1}$ (green), respectively, with the associated shallow dips at around $b=4.3$ \AA~ and $b=3.9$ \AA.
Also, we see shoulders around $b=3.9$ \AA~at higher energies ($E_\text{c}>3000$ cm$^{-1}$).
These might be a manifestation of the high-$b$ peak ($l$-type rainbow) in the QCT results. 
From the resulting trajectories, it is possible to directly confirm the occurrence of $l$-type rainbows from the correlation between the scattering angle and the impact parameter as shown in \cref{fig:deflection}. 
At $E_\text{c}=500$  and $1000$ cm$^{-1}$, we see a clear signature of an $l$-type rainbow with a maximum at around $b_\text{r}=4.2$ \AA, where a subset of trajectories give rise to the same scattering angle, $\theta_\text{r} \sim 10 ^\circ$-$20 ^\circ$. 
The fact that the same scattering angles are reached by trajectories with different impact parameters implies that in quantum mechanics these different scattering paths could constructively or destructively interfere, leading to the rainbow signature.
We also see the glory impact parameter at around $b_\mathrm{g}=3.6$ \AA\  arising from a balance between the repulsive and attractive forces resulting in a scattering angle of $\sim 0^\circ$. 
At higher energies, the maximum arising from the rainbow is not clear due to the small magnitude of $\theta$ above $b_\text{g}$ as seen in~\cref{fig:deflection} (c) and (d). 

\begin{figure}[b]
\begin{center}
\includegraphics[scale=0.4]{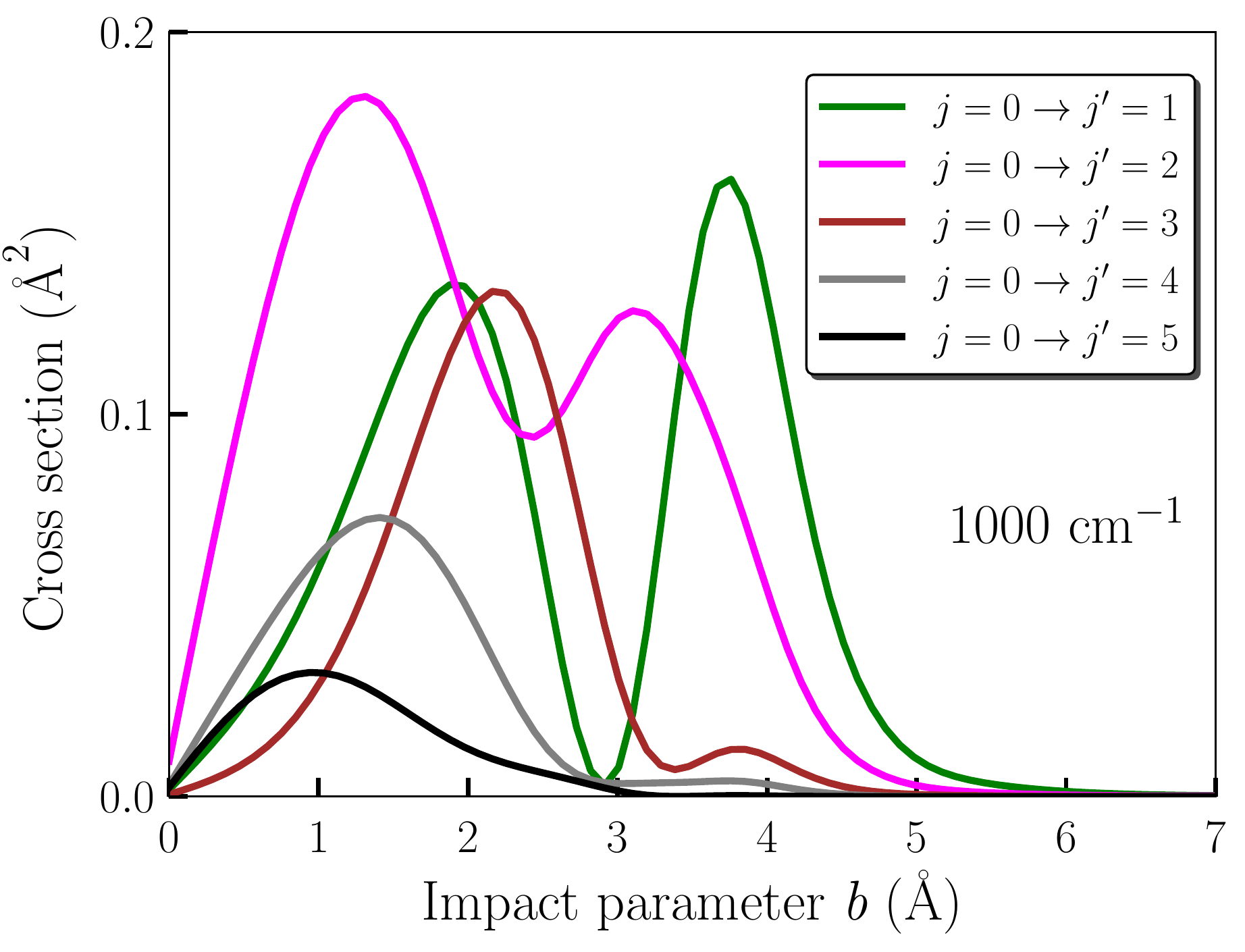}
\end{center}
\vspace{-5mm}
\caption{
Partial rotational inelastic cross section as a function of impact parameter $b$ for rotational excitations out of the ground state HCl ($v=0, j=0$) to ($v'=0, j'$) ($j'=1$ to $5$) in collisions with {\it para}-H$_2$ ($ j=0$) from CC calculations at $E_\text{c}=1000$ cm$^{-1}$.
}
\label{fig:high_j}
\end{figure}

While we see a clear signature of an $l$-type rainbow in the correlation diagram in \cref{fig:deflection}, it is worth noting the discrepancies between CC and QCT results in panels (a) vs (b) in \cref{fig:b-ICS}, including the values of glory and rainbow impact parameters extracted from \cref{fig:deflection}.
For rotational excitations in Ar+NO collisions, Aoiz and coworkers~\cite{Aoiz_2003_Ar-NO} 
demonstrated excellent agreement of the opacity functions derived from quantum mechanical (QM) and QCT calculations, including the locations of minimum and maximum.
Similar good agreement between QM and QCT results in the partial ICS or opacity functions were reported for Xe+NO and Cl+H$_2$.~\cite{Aoiz_2011_Cl-H2,Aoiz_2012_Xe-NO}
In these systems, there is a region of $b$ ($\sim b_\text{g}$) in which only very few trajectories are observed, leading to the discontinuity in the distribution of trajectories.
The discontinuity is the origin of the minimum that becomes almost zero in the partial ICS or opacity function in their QCT results. 
On the contrary, we do not observe such discontinuity around $b_\text{g}$ in our  trajectory results (\cref{fig:deflection}).
The limitation of QCT calculation to reproduce QM result was also reported for Kr+NO, and the discrepancy is attributed to the sensitivity of the interference effect present in the quantum results.~\cite{Aoiz_2014_Kr-NO}  
For Ar+HF, a classical mechanical treatment did not reproduce a minimum structure in the opacity function obtained with a semiclassical treatment that includes the interference effect.~\cite{Barrett_semi}
Strictly, the origin of rainbow signature is quantum interference while the existence of the rainbows is conveniently explained by the singularities observed in classical mechanics. 
Thus, the observed discrepancy between QM and QCT may not be altogether surprising.
Further systematic studies are required to isolate systems for which classical and quasi-classical calculations are not effective in reproducing specific rainbow signatures.

\begin{figure}[b]
\begin{center}
\includegraphics[scale=0.4]{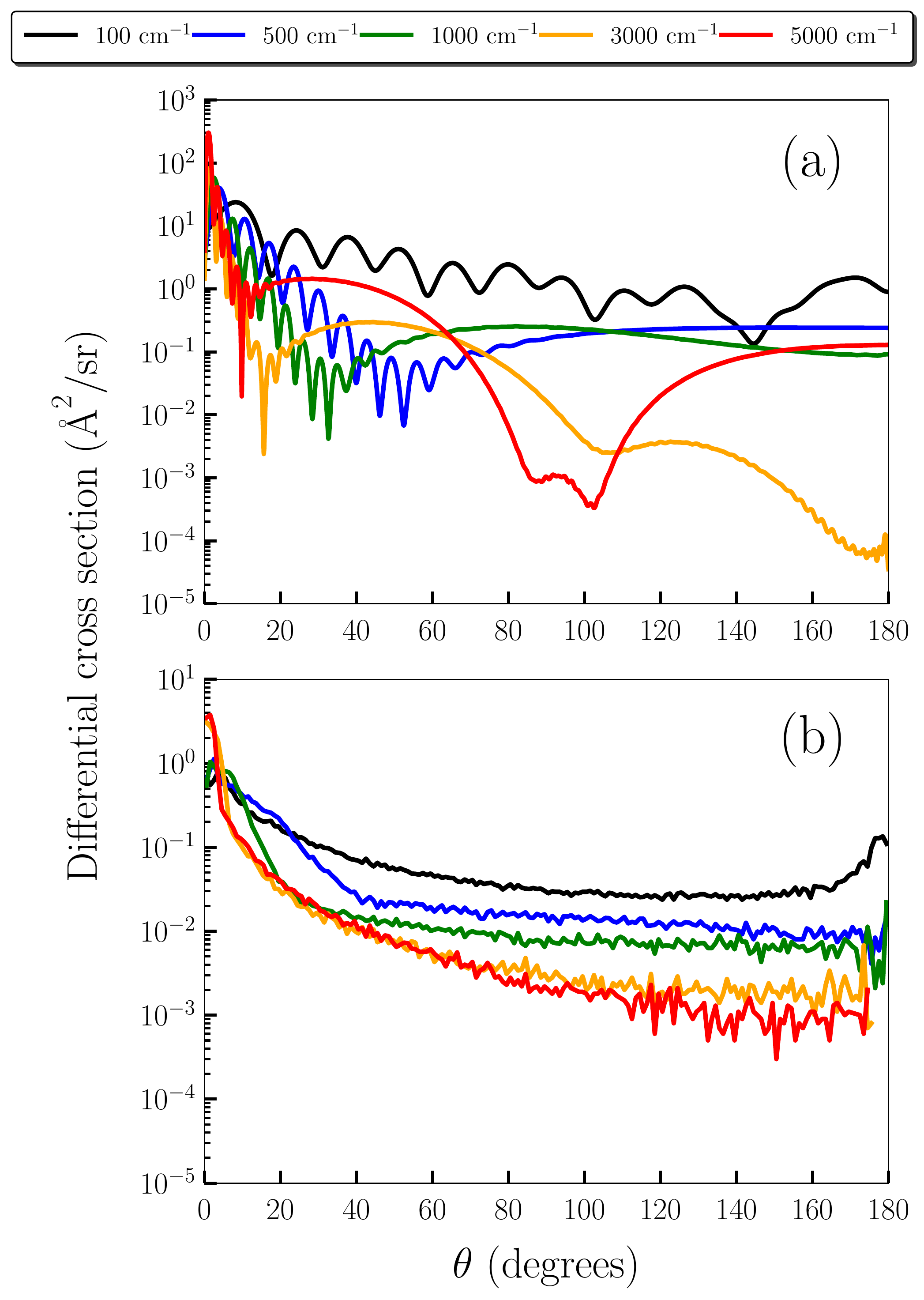}
\end{center}
\vspace{-5mm}
\caption{
Differential cross section as a function of the scattering angle for HCl ($v=0, j=0$) to ($v'=0, j'=1$) in collisions with {\it para}-H$_2$ ($ j=0$).
(a) CC and (b) QCT.  
}
\label{fig:DCS}
\end{figure}

Next, we  examine whether the nature of the high-$b$ peak observed in the CC calculation (\cref{fig:b-ICS}~(a)) is consistent with an $l$-type rainbow. 
Since $l$-type rainbows for rotational excitation arise from anisotropy in the long-range attractive part of the potential, it would become difficult to observe $l$-type rainbows in the rotational transition processes that accompany large changes in rotational levels.~\cite{Aoiz_2011_Cl-H2}
To illustratemu this, we show  results for rotational excitations to higher rotational states at $E_\text{c}=1000$ cm$^{-1}$ (\cref{fig:high_j}).
Compared to $|\Delta j|=1$ (green), we see a gradual suppression of  contributions from  $\sim b>3$ \AA\ with increasing $|\Delta j|$. 
In particular, for $|\Delta j|=4$ (black), the contribution from high $b$ vanishes. 
This result combined with the discussions of the energy dependence of the ICS in \cref{fig:b-ICS}~(a) and the QCT trajectories distribution in \cref{fig:deflection} support the conclusion that the high-$b$ peak observed in the  $|\Delta j|=1$ cross sections is an $l$-type rainbow.  

We have confirmed that long-range attractive forces give rise to a modest well in the potential and lead to the  high-$b$ peak ($l$-type rainbow) in the partial ICS (\cref{fig:b-ICS}~(a)).
On the other hand, the dynamics at low-$b$ is significantly influenced by the inner repulsive region of the interaction potential, thus the peaks observed at low-$b$ could be ascribed to a rotational rainbow.   
One typical way to examine the occurrence of the rotational rainbow is to explore the DCS and its final state $j'$ dependence for a given scattering angle,~\cite{Schinke_l_Li+-H2,Schinke_1981,GianturcoPalma,Rev_KorschErnesti} thus we move on to the analysis of the DCS.

In \cref{fig:DCS} we show the energy dependence of the DCS for $j=0$ to $j'=1$ transition in HCl from CC (a) and QCT (b) calculations. 
As discussed earlier, high-$b$ and low-$b$ collisions correspond to small (forward) and large (backward) scattering angle ($\theta$), respectively.
Therefore, there exists a correspondence between \cref{fig:b-ICS} and \cref{fig:DCS}, although different values of $b$ can result in the same scattering angle $\theta$  even in the classical limit.
Indeed, in 
\cref{fig:DCS}~(a),
we observe two broad peaks in the DCS above $\theta=30 ^\circ$ corresponding to the two peaks in low-$b$ region in \cref{fig:b-ICS}~(a) above $E_\text{c}=3000$. 
As the collision energy is decreased below $E_\text{c}=3000$, the two broad peaks completely vanish at $E_\text{c}=500$ cm$^{-1}$, consistent with the energy dependence of the low-$b$ peak in \cref{fig:b-ICS}~(a).

We can also observe signatures of the  $l$-type rainbow at small angle scattering  in the DCS. 
Below the inflection point, the cross section increases rapidly with a decrease of $\theta$. 
The inflection point is an indicator of the region of the $l$-type rainbow. 
Indeed, the energy dependence of the position of the inflection point is consistent with the minimum that separates the low-$b$ and high-$b$ regions in \cref{fig:b-ICS}~(a). 
However, as we can see, diffraction pattern makes it difficult to identify the positions of the maxima of $l$-type rainbow and its secondary (supernumerary) rainbows. 

\Cref{fig:DCS}~(b) indicates the capability of QCT in describing the behavior of DCS.  
Similar to the discussion of \cref{fig:b-ICS}, the DCS from QCT fails to reproduce the specific details of the quantum DCS although the overall behavior is qualitatively reproduced. 
The broad peaks in the large $\theta$ region or the detailed structures in the low-$b$ region in \cref{fig:b-ICS}~(b) are not reproduced in the QCT results.
On the other hand, we see a  change in the behavior of the DCSs at lower $\theta$ corresponding to the inflection points in \cref{fig:DCS}~(a), the boundary between low-$b$ and high-$b$ regions, while the signature of $l$-type rainbow is not evident or correct.
Features of $l$-type rainbows in classical mechanics are expected to appear as an abrupt change of the cross section separating the region of $\theta_\text{r}$ into bright side ($\theta<\theta_\text{r}$) and dark side ($\theta>\theta_\text{r}$).~\cite{Schinke_1982,Aoiz_2003_Ar-NO} We find a slight change in the behavior of the small peaks at very low $\theta$ ($< 5 ^\circ$) at  higher collision energies in \cref{fig:DCS}~(b).

\begin{figure}[tb!]
\begin{center}
\includegraphics[scale=0.4]{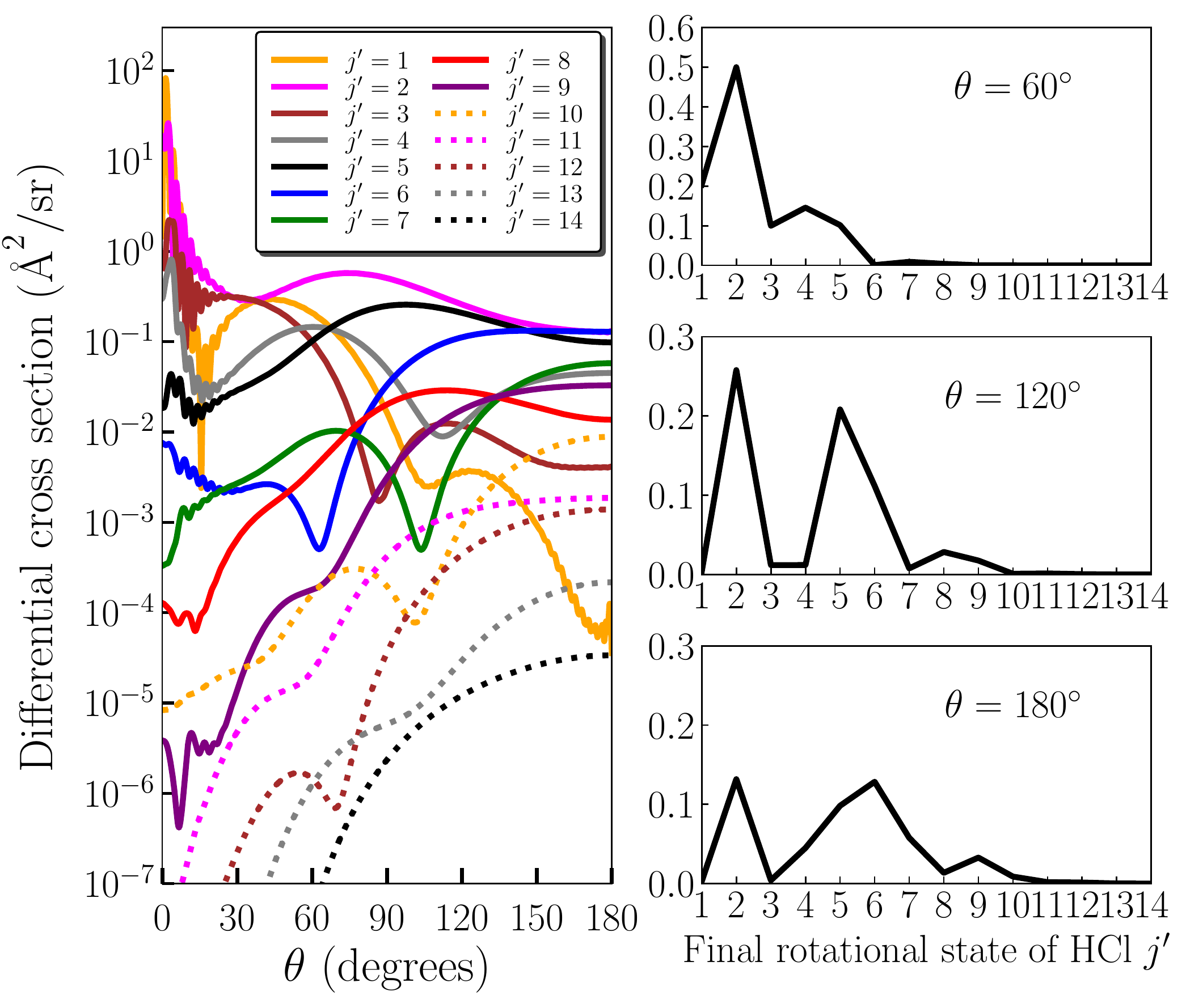}
\end{center}
\vspace{-5mm}
\caption{
Left panel: state-to-state differential cross section as a function of scattering angle for rotational excitations out of the ground state HCl ($v=0, j=0$) to ($v'=0, j'$) ($j'=1$ to $14$) in collisions with {\it para}-H$_2$ ($ j=0$) by CC calculations at $E_\text{c}=3000$ cm$^{-1}$.
Right panels: differential cross section as a function of final rotational state $j'$ for selected values of the scattering angle. 
}
\label{fig:DCSexcite}
\end{figure}

To further examine the rainbow signatures in the DCS from quantum calculations, in the left panel of \cref{fig:DCSexcite}, we show the DCS at $E_\text{c}=3000$ cm$^{-1}$ for rotational excitations out of the ground state of HCl  to all energetically accessible rotational states ($v'=0, j'=1$ to $14$) in collisions with {\it para}-H$_2$.
The behavior and magnitude of DCS at small $\theta$ is drastically changed at $j'=5$ (black curve) due to the disappearance of $l$-type rainbow, which is consistent with the discussion of the partial ICS in \cref{{fig:high_j}}.
We observe an overall decreasing trend of the cross sections with increasing $j'$. 
However, it is possible to find some exception to this trend suggesting more nuanced nature of the  $j'$ dependence of the DCS in identifying features of rotational rainbows.
In particular for $j' \geq 5$, we observe a  gradual shift of the broader peaks to the higher $\theta$ region with increasing $j'$. 
This trend is consistent with  signatures of a rotational rainbow.~\cite{LevinBernstein,Schinke_1982}
The right panels of \cref{{fig:DCSexcite}}  depict DCS as a function of $j'$ for $\theta=60, 120$ and $180 ^\circ$. 
The results for $\theta=120$ and $180 ^\circ$ mainly reveal the $j'$ dependence of the broader peaks observed in the DCS, related to the low-$b$ peak in the partial ICSs.   
The appearance of multiple peaks with respect to $j'$ seems to be consistent with signatures of rotational rainbows.
However, the distributions of the peaks are significantly different from typical $j'$-distributions for heteronuclear target molecules that feature multiple peaks with two distinct maxima~\cite{exp_hetero,hetero_Schinke_1980,expcalc_hetero}. 
In particular, the $j'$ distributions in \cref{fig:DCSexcite} is characterized by a peak at  low $j'$ ($\sim j'=2$) and the absence of an intense peak at high $j'$.
Very similar signatures were reported in atom+molecule systems with a large asymmetric mass for the diatomic molecule, He+HF\,~\cite{GianturcoPalma} and Li$^{+}$+BeF\,~\cite{Nakamura}. 
The overall trend is robust with respect to the collision energy. Indeed, in \cref{{fig:high_j}}, we can see that the excitation to $j'=2$ results in the largest (total) cross section at 1000 cm$^{-1}$ (see also SM).

\section{Conclusion}
\label{sec:Conclusion}

In this paper, we performed full-dimensional quantum mechanical scattering calculations using the close-coupling formalism for rotationally inelastic scattering of HCl ($v=0,j=0$) in collisions with ground state {\it para}-H$_2$ on a globally accurate full-dimensional  {\it ab initio} potential energy surface.
Our results for rotational excitations of HCl ($j=0 \to j'$) at collision energies ranging from 100 to 6000 cm$^{-1}$ show signatures of rainbow scattering in both state-to-state partial integral and differential cross sections. 

For the $j=0 \to j'=1$ excitation, signatures of $l$-type rainbow are observed in the impact parameter dependence of the partial integral cross sections as a pronounced peak in the higher impact parameter region.
This peak  is clearly separated from a peak at a lower $b$ with a minimum where the cross section becomes almost zero at around $b=3.0$ \AA.
Features of the high-$b$ peak are examined in terms of the collision energy and final rotational quantum numbers, confirming the appearance of $l$-type rainbow. 
In addition, we demonstrated that $l$-type rainbow does not occur for high rotational excitations of HCl since the anisotropy of the long-range attractive force is not sufficiently strong to generate the required torque to  excite high rotational states. 
The correlation diagram of impact parameter and scattering angle obtained from quasi-classical trajectory calculations also qualitatively supports the presence of an $l$-type rainbow. 
These analyses indicate the coexistence of distinctive dynamical regimes for HCl rotational transition driven by the short-range repulsive and long-range attractive  forces whose relative importance depends on the collision energy and final rotational states.
We have also identified characteristic multi-peak structure in the final rotational state dependence of  the differential  cross section that are consistent with previously studied atom+heteronuclear molecule systems with a large asymmetric mass for the molecules (composed of light and heavy atoms).

\begin{acknowledgments}
This work was supported in part by ARO MURI grant No. W911NF-19-1-0283 (H.G. and N.B.) and NSF grant No. PHY-1806334 (N.B.). 
The calculations at UNM were performed at the Center for Advanced Research Computing (CARC).
\\ 
\end{acknowledgments}

\section*{Data availability statement}
The data that support the findings of this study are available within the article and its supplementary material.

\bibliography{cite}



\clearpage
\onecolumngrid
\vspace{\columnsep}

\newcolumntype{Y}{>{\centering\arraybackslash}X}
\newcolumntype{Z}{>{\raggedleft\arraybackslash}X}

\setcounter{figure}{0}
\setcounter{equation}{0}
\setcounter{page}{1}

\renewcommand{\thepage}{S\arabic{page}}
\renewcommand{\thefigure}{S\arabic{figure}}
\renewcommand{\theequation}{S\arabic{equation}}

\onecolumngrid

\begin{center}
	\textbf{\huge Supplementary Material}
\end{center}

\begin{center}
\text{ \Large 
Rainbow scattering in rotationally inelastic collision of HCl and H$_2$
}
\end{center}

\begin{center}
\text{ Masato Morita, Junxiang Zuo, Hua Guo and Naduvalath Balakrishnan} 
\end{center}

\text{}


\setcounter{section}{1}
\begin{center}
\textbf{ \large
\thesection{\label{sec:1}.	Partial integral cross section}
}
\end{center}

In Fig.~1 of the main text, partial integral cross section (ICS) obtained with quantum mechanical close-coupling (CC) calculation is displayed as a function of the incoming partial wave $l$ for the rotational excitation of HCl by  {\it para}-H$_2$, namely HCl ($v=0$, $j=0$) + H$_2$ ($j=0$)  $\to$ HCl ($v'=0$, $j'=1$) + H$_2$ ($j'=0$). 
Since the initial rotational states are the ground states ($j=0$) for both HCl and H$_2$, the total angular momentum $J$ of the collision complex is equal to $l$. 
On the other hand, due to the conservation of $J$, the possible outgoing partial wave $l'$ for this process is  $l'=J-1$ and $J+1$, thus $l'=l-1$ and $l+1$. 
In other words, each $l\,(=J)$ component of the partial wave cross section in Fig.~1 is given as the sum of two outgoing components.   

\begin{figure}[b!]
\begin{center}
\includegraphics[scale=0.4]{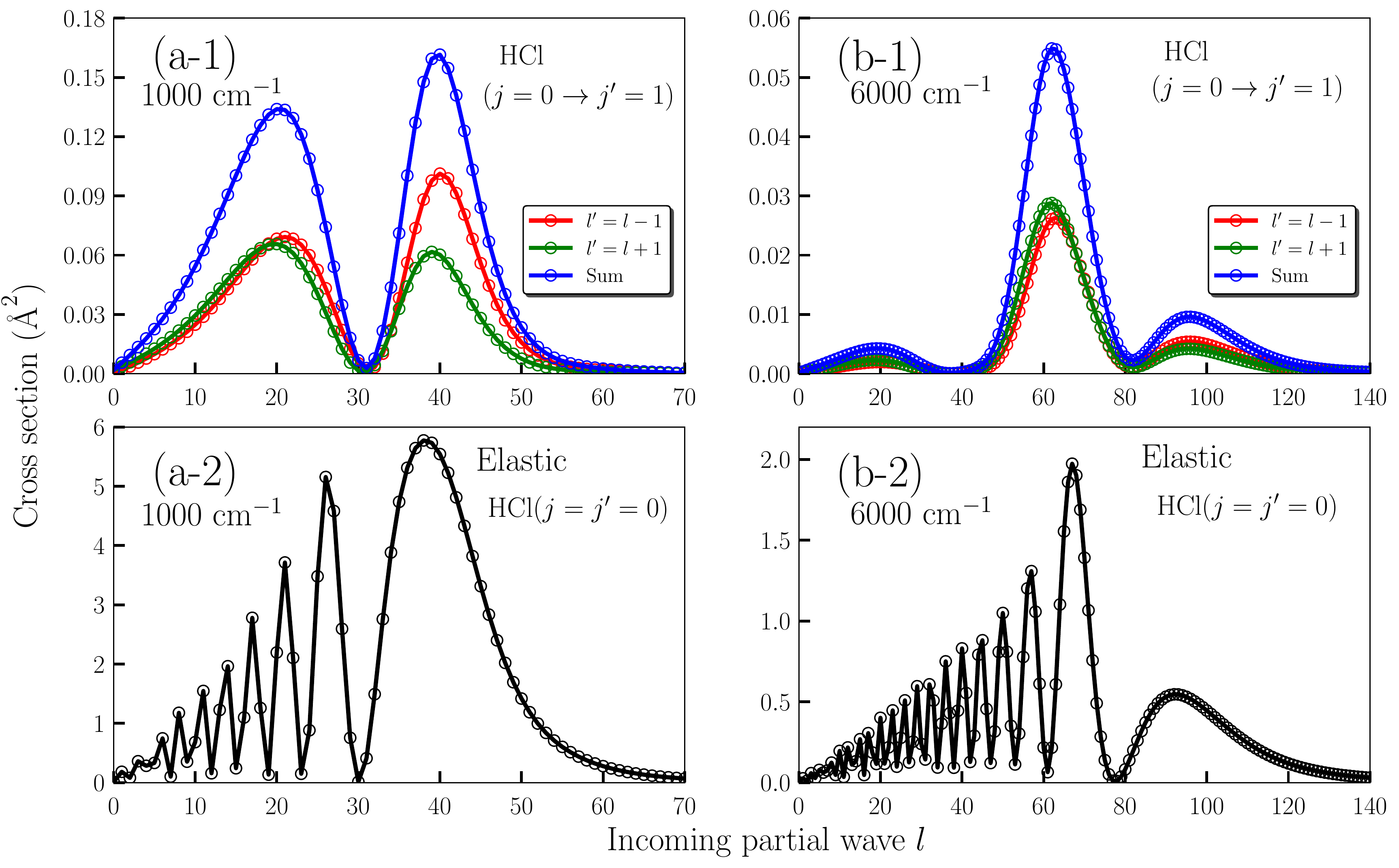}
\end{center}
\caption{ 
(a-1) and (b-1): partial integral cross section (blue) for the rotationally inelastic collisions of HCl ($j=0 \to j'=1$) in collisions with {\it para}-H$_2$ ($j=0$), HCl ($v=0, j=0$) + H$_2$ ($ j=0$) $\to$ HCl ($v'=0, j'=1$) + H$_2$ ($ j'=0$), at $E_\text{c}=1000$ and $6000$ cm$^{-1}$, respectively.
Decomposition into $l'=l-1$ (red) and $l'=l+1$ (green) components is also displayed. 
(a-2) and (b-2): partial integral cross section (black) for the elastic collision between HCl ($j=0$) and {\it para}-H$_2$ ($j=0$), HCl ($v=0, j=0$) + H$_2$ ($ j=0$) $\to$ HCl ($v'=0, j'=0$) + H$_2$ ($j'=0$).
} 
\label{fig:SI_1}
\end{figure}

In \cref{fig:SI_1} (a-1) and (b-1), we show the $l'=l-1$ (red) and $l'=l+1$ (green) components of the partial ICS (blue) as a function of $l$ at the collision energies of $E_\text{c}=1000$ and $6000$ cm$^{-1}$. 
As we can see, the $l'=l-1$ and $l'=l+1$ components exhibit an almost identical $l$-dependence  except for the absolute values. 
Thus, the minima and maxima occur at almost the same $l$ in all curves in the figures. 
In particular, the first minimum on the right, separating the low-$l$ and high-$l$ (at around $l\sim30$ at $E_\text{c}=1000$ and $l\sim80$ at $E_\text{c}=6000$), is linked to the elastic cross section for HCl ($v=0$, $j=0$) + H$_2$ ($j=0$) $\to$ HCl ($v'=0$, $j'=0$) + H$_2$ ($j'=0$) as shown in \cref{fig:SI_1} (a-2) and (b-2). 
It is seen that the  minimum that separates the broad peak in the high $l$ region (a-2)/(b-2)  is nearly the same as in (a-1)/(b-1), though slightly displaced in $l$. 
Surely, this relation is one of the proofs of the $l$-type rainbow character of the high-$l$ (high-$b$) peak in the rotational excitation of HCl in (a-1) and (b-1). 
In other words, to correctly describe the behavior of the partial ICS at around the minimum separating the high-$l$ and low-$l$ regions, it is necessary to reflect the information of the phase shift during scattering in the long-range part of the potential  regardless of the rotational transitions of molecules. 

\newpage

\setcounter{section}{2}
\begin{center}
\textbf{ \large
\thesection{\label{sec:2}.	Isotopic effect}
}
\end{center}

\begin{figure}[h!]
\begin{center}
\includegraphics[scale=0.39]{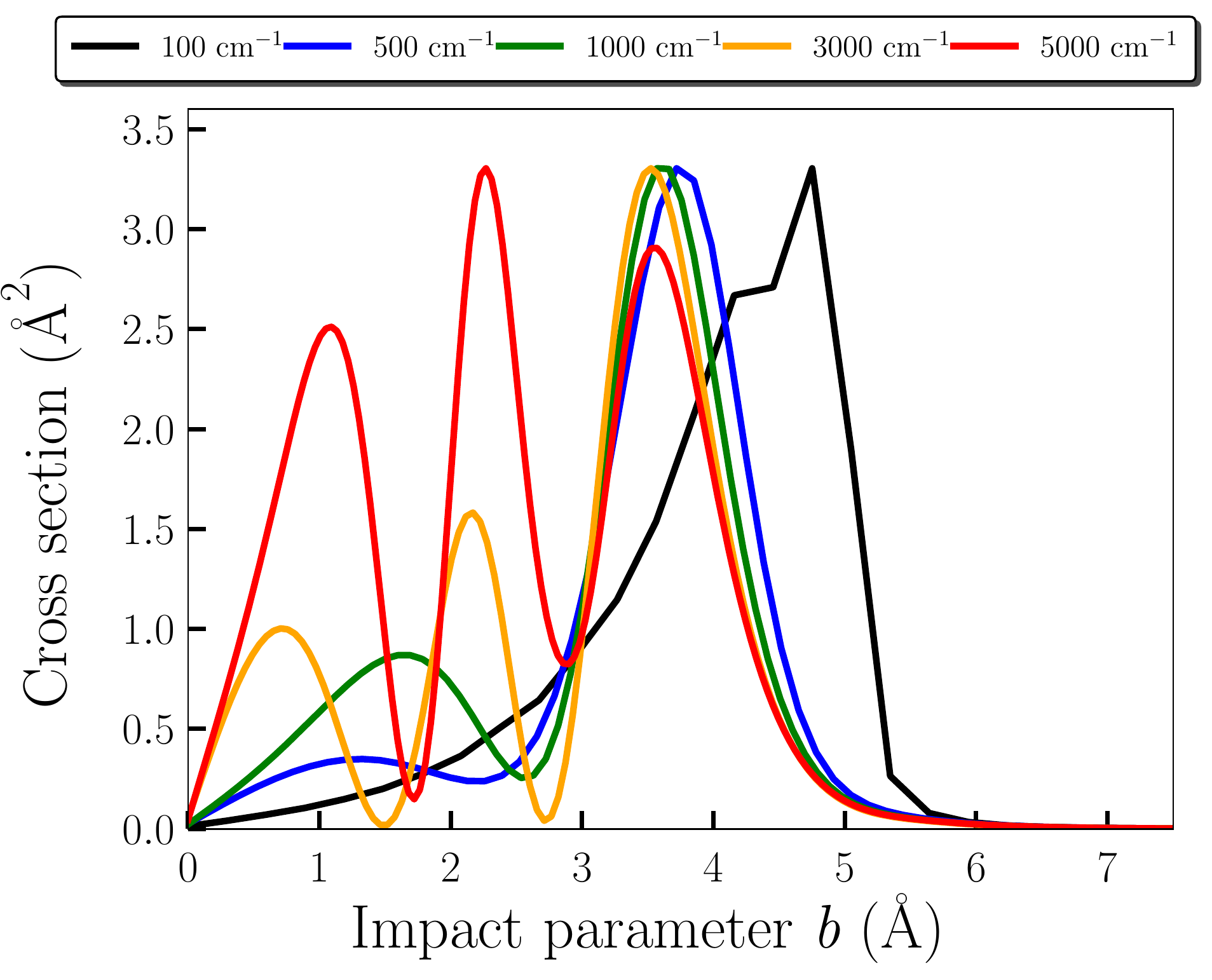}
\end{center}
\caption{ 
Partial integral cross section for rotationally inelastic transitions in DCl ($j=0 \to j'=1$)  by collisions {\it para}-H$_2$ ($j=0$), DCl ($v=0, j=0$) + H$_2$ ($ j=0$) $\to$ DCl ($v'=0, j'=1$) + H$_2$ ($j'=0$).
} 
\label{fig:SI_2}
\end{figure}

\Cref{fig:SI_2} shows the partial ICS for the rotationally inelastic collisions of DCl ($j=0 \to j'=1$) in collisions with {\it para}-H$_2$ ($j=0$). 
In comparison with Fig.~2(a) in the main text, we see similar but slight distinct multi-modal structures as a function of $b$ including the positions of peaks and minima. 
Since the interaction potential is unaffected by the substitution of the H atom by the D atom, the deviations arise from the difference in  the reduced mass and hence the de-Bloglie wave length.  

\setcounter{section}{3}
\begin{center}
\textbf{ \large
\thesection{\label{sec:3}.	Excitation from a rotationally excited state (HCl: $j=1 \to j'=2$) }
}
\end{center}

\begin{figure}[h!]
\begin{center}
\includegraphics[scale=0.39]{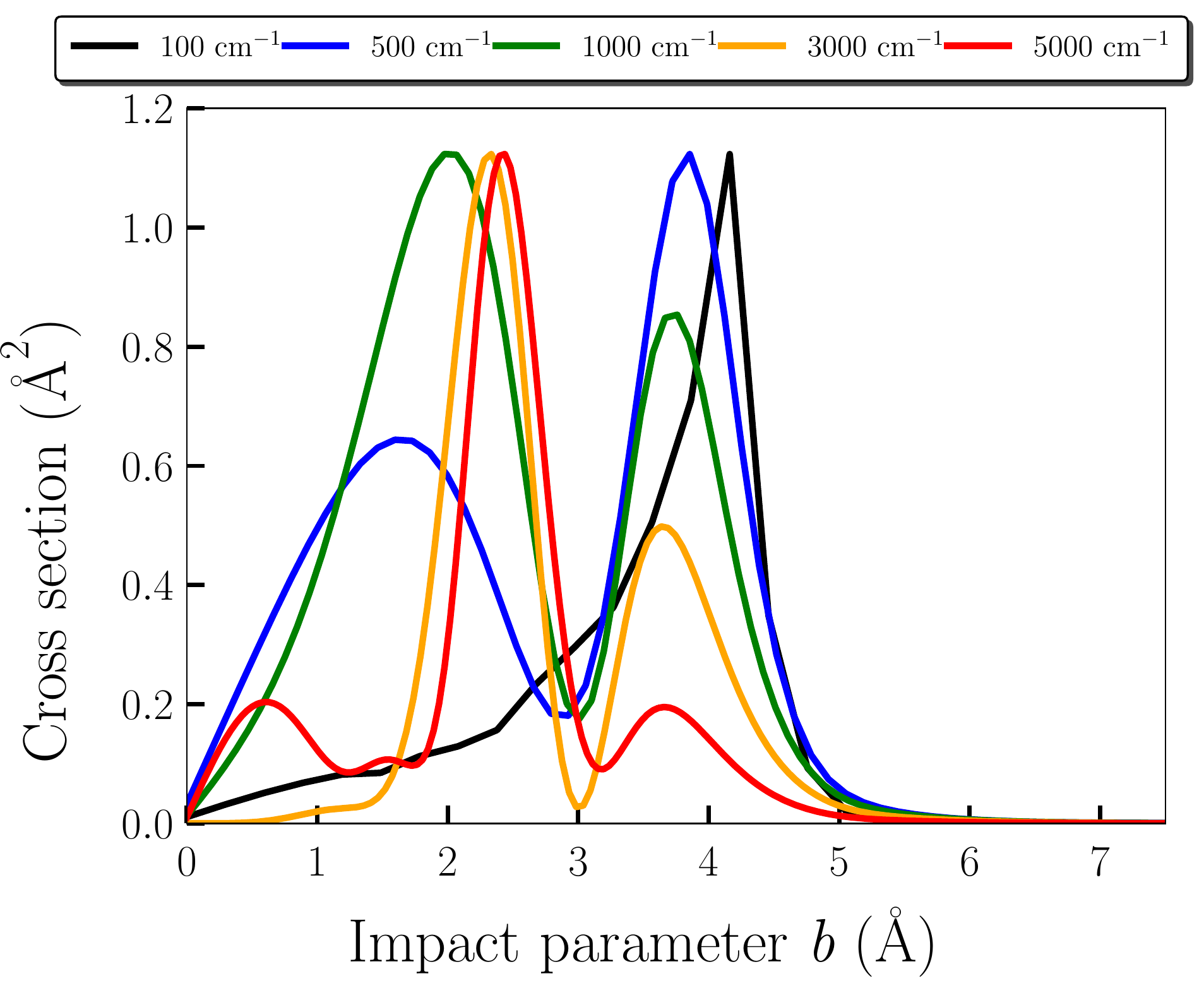}
\end{center}
\caption{ 
Partial integral cross section for the rotationally inelastic scattering of HCl ($j=1 \to j'=2$) in collisions with {\it para}-H$_2$ ($j=0$), HCl ($v=0, j=1$) + H$_2$ ($j=0$) $\to$ HCl ($v'=0, j'=2$) + H$_2$ ($j'=0$).
} 
\label{fig:SI_3}
\end{figure}

The partial ICS from the rotationally excited state ($j=1$) to $j'=2$ in \cref{fig:SI_3} shows similar behavior as in Fig.~2 (a) of the main text ($j=0 \to j'=1$).
Also, we observe similar dependence for large $j'$ as in the main text ($j=0$).  
This implies that $\Delta j = j'-j$ is  more suitable as a fundamental variable than $j'$ itself to characterize (rainbow) scattering \cite{Bowman_1979,SchinkeMuller}. 

\newpage

\setcounter{section}{4}
\begin{center}
\textbf{ \large
\thesection{\label{sec:4}.	$j'$-dependence of the differential cross section}
}
\end{center}

\begin{figure}[h!]
\begin{center}
\includegraphics[scale=0.42]{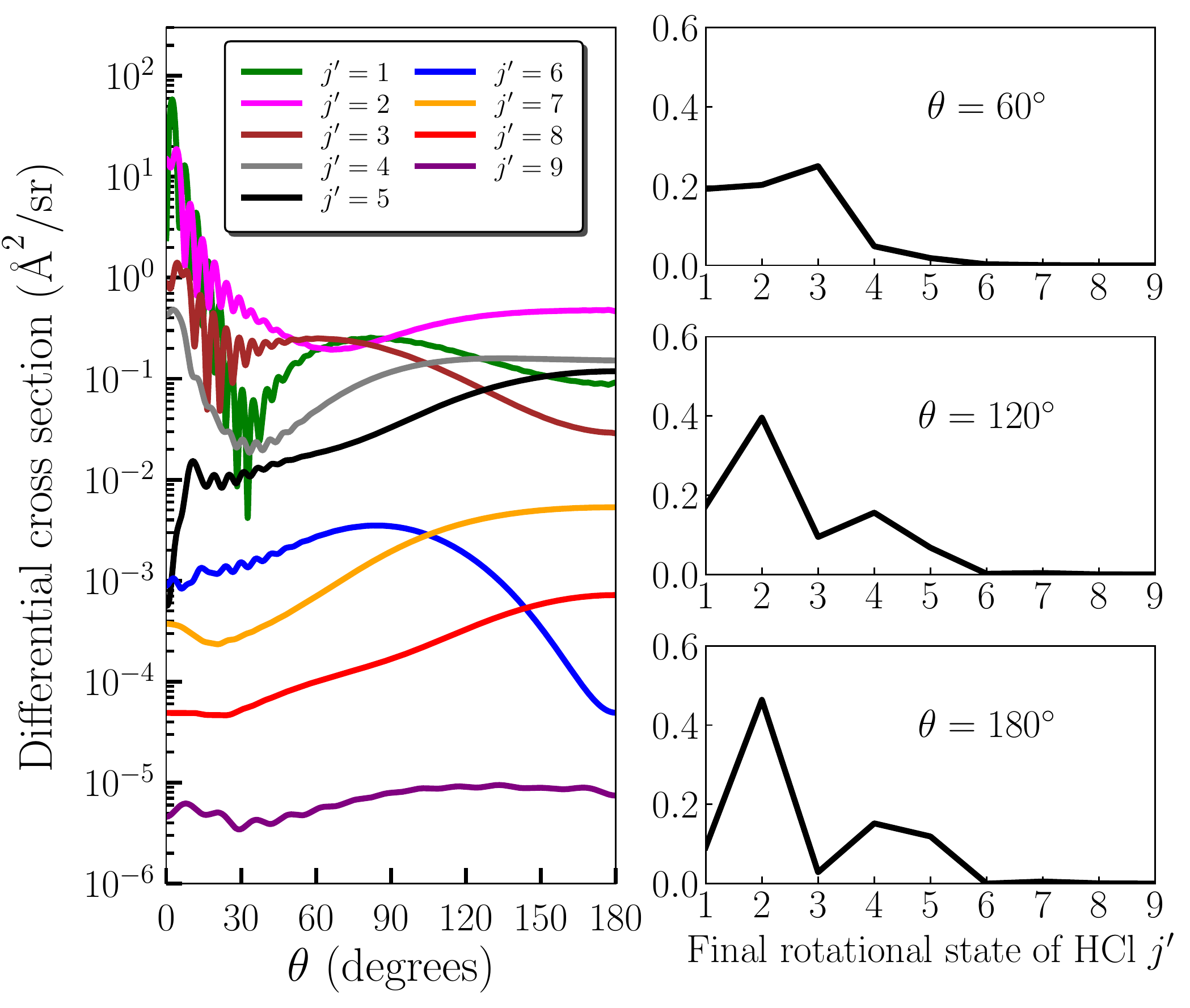}
\end{center}
\caption{ 
Left panel: state-to-state differential cross section as a function of the scattering angle for rotational excitations out of the ground state of HCl ($v=0, j=0$) to ($v'=0, j'$) ($j'=1$ to $9$) in collisions with {\it para}-H$_2$ ($ j=0$) from CC calculations at $E_\text{c}=1000$ cm$^{-1}$.
Right panels: differential cross section as a function of final rotational state $j'$ at selected values of the scattering angle. 
} 
\label{fig:SI_4}
\end{figure}

\begin{figure}[h!]
\begin{center}
\includegraphics[scale=0.42]{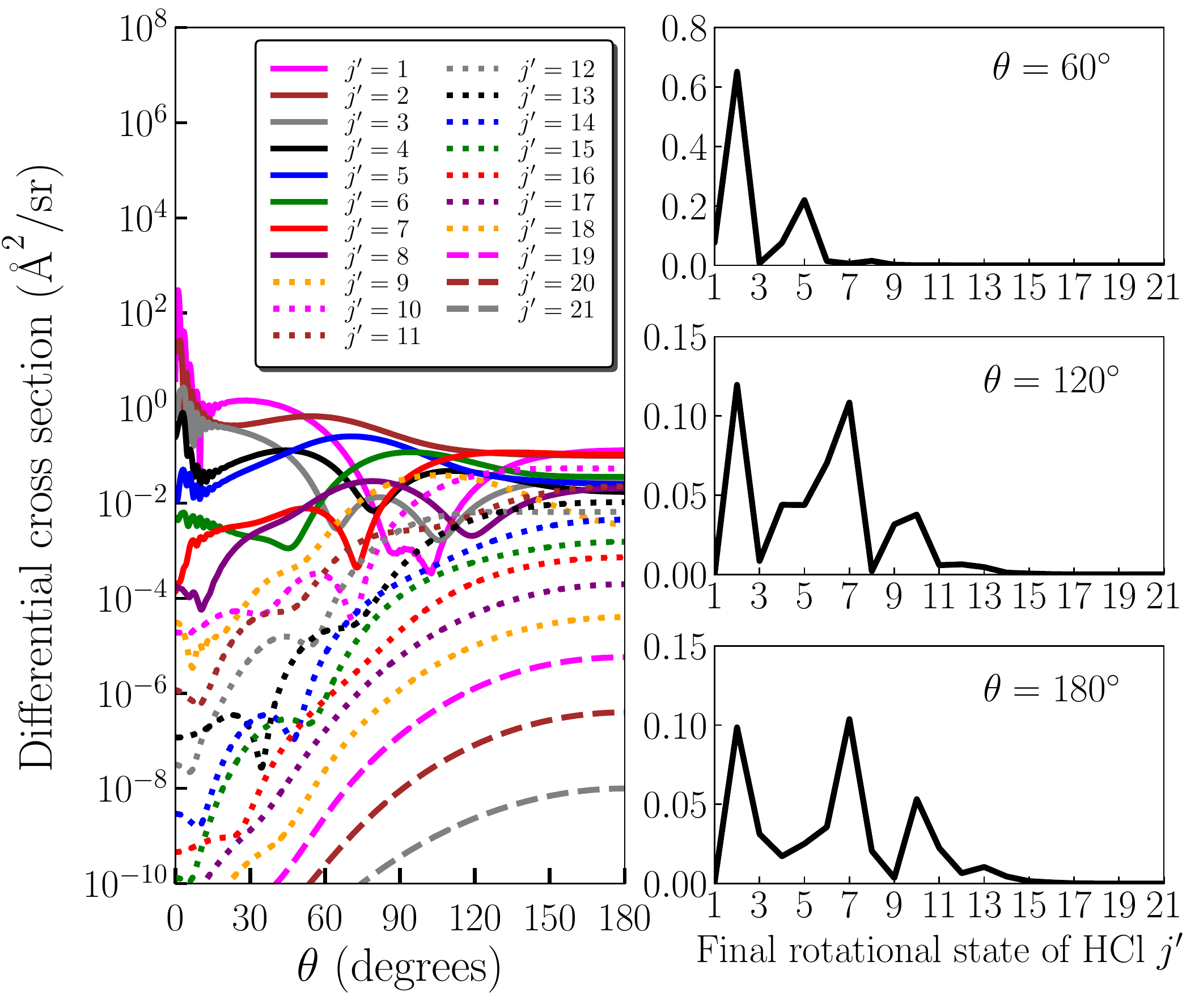}
\end{center}
\caption{ 
Similar results as \cref{fig:SI_4} but for $j'=1$ to $21$ in HCl at $E_\text{c}=5000$ cm$^{-1}$.
} 
\label{fig:SI_5}
\end{figure}

In Fig.~6 of the main text, we show the behavior of the DCS with respect to the final rotational quantum number $j'$ at a collision energy of $E_\text{c}=3000$ cm$^{-1}$.  
We illustrate similar behavior at  collision energies of $E_\text{c}=1000$ and 5000 cm$^{-1}$ in \cref{fig:SI_4,fig:SI_5} although the peak structures as a function of $j'$ become more clear with increasing  collision energy. 

\end{document}